\documentclass[a4paper,11pt]{article}
\usepackage{pos}

\title{Sensitivity of DUNE to low energy physics searches}

\author*[a]{C. Cuesta}
\author{on behalf of DUNE collaboration }

\affiliation[a]{ Centro de Investigaciones Energé$ticas, Medioambientales y Tecnol$ógicas,
 CIEMAT, \\ 28040, Madrid, Spain}

\emailAdd{clara.cuesta@ciemat.es}

\abstract{The Deep Underground Neutrino Experiment (DUNE), a next-generation long-baseline neutrino oscillation experiment, is a powerful tool to perform low energy physics searches. DUNE will be uniquely sensitive to the electron-neutrino-flavour component of the burst of neutrinos expected from the next Galactic core-collapse supernova, and also capable of detecting solar neutrinos. DUNE will have four modules of 70-kton liquid argon mass in total, placed 1.5 km underground at the Sanford Underground Research Facility in the USA. These modules are being designed exploiting different liquid argon time projection chamber technologies and based on the physics requirements that take into account the particularities of the low energy physics searches.}

\FullConference{%
  41st International Conference on High Energy physics - ICHEP2022\\
  July 6 - 13, 2022\\
  Bologna, Italy
}

\begin{document}
\maketitle
\section{The Deep Underground Neutrino Experiment}

The Deep Underground Neutrino Experiment (DUNE) is a next-generation long-baseline neutrino oscillation experiment with a primary physics goal of observing neutrino and antineutrino oscillation patterns to precisely measure the parameters governing long-baseline neutrino oscillation in a single experiment, and to test the three-flavor paradigm~\cite{DUNE_LBL}. DUNE is also uniquely sensitive to low energy neutrinos such as electron neutrinos from a galactic supernova burst (SNB)~\cite{DUNE_SNB}, and to a broad range of physics beyond the Standard Model (BSM)~\cite{DUNE_BSM}, including nucleon decays. 

DUNE will consist of a near detector placed at Fermilab close to the production point of the muon neutrino beam of the Long-Baseline Neutrino Facility (LBNF), and four 17\,kt liquid argon time projection chambers (LArTPCs) as far detector (FD) in the Sanford Underground Research Facility (SURF) at 4300\,m.w.e. depth at 1300\,km from Fermilab. 

DUNE is anticipated to begin collecting physics data with Phase I, an initial experiment configuration consisting of two FD modules and a minimal suite of near detector components, with a 1.2\,MW proton beam. The Phase II upgrades necessary to achieve DUNE's physics goals are: addition of FD modules three and four for a total FD fiducial mass of at least 40\,kt, upgrade of the proton beam power to 2.4\,MW and of the near detector.

\section{The DUNE Far Detector}

The DUNE FD will be sensitive to low energy neutrinos from astrophysical sources such as a SNB and the Sun. The FD LArTPC technology provides good energy resolution, full particle reconstruction with very high quality tracking, and energy thresholds as low as a few MeV may be possible. 

In these detectors, the ionization charge is drifted by an electric field towards the anode where the charge is collected. Using the time arrival of the charge at the readout planes, a three-dimensional track reconstruction is possible.  Particles are identified by the rate of energy loss along the track.  The Ar scintillation light is also detected enabling fast timing of signals and event localization inside the detector.

Different LAr technologies are being considered for the DUNE FD: the first module will employ the single-phase horizontal-drift (HD) LArTPC technology~\cite{tdr_v4}, where the drift of 3.5\,m is horizontal with wrapped-wire readout including two induction and one charge collection anode planes \cite{tdr_v4}; and the second module will employ the single-phase vertical-drift (VD) LArTPC technology where the drift is vertical over 6\,m. For low-energy signals, such as solar and SNB neutrinos, the VD design provides opportunities to improve the physics performance, primarily due to improved photon detection.

\section{Low energy events in DUNE}

The few-MeV low energy regime is of particular interest for the detection of the burst of neutrinos from a galactic core-collapse supernova, which has been the primary focus of DUNE low-energy sensitivity studies, but DUNE will also have sensitivity to neutrinos from other astrophysical sources, including solar neutrinos.

\begin{figure}[htp]
    \centering
    \includegraphics[width=0.6\textwidth]{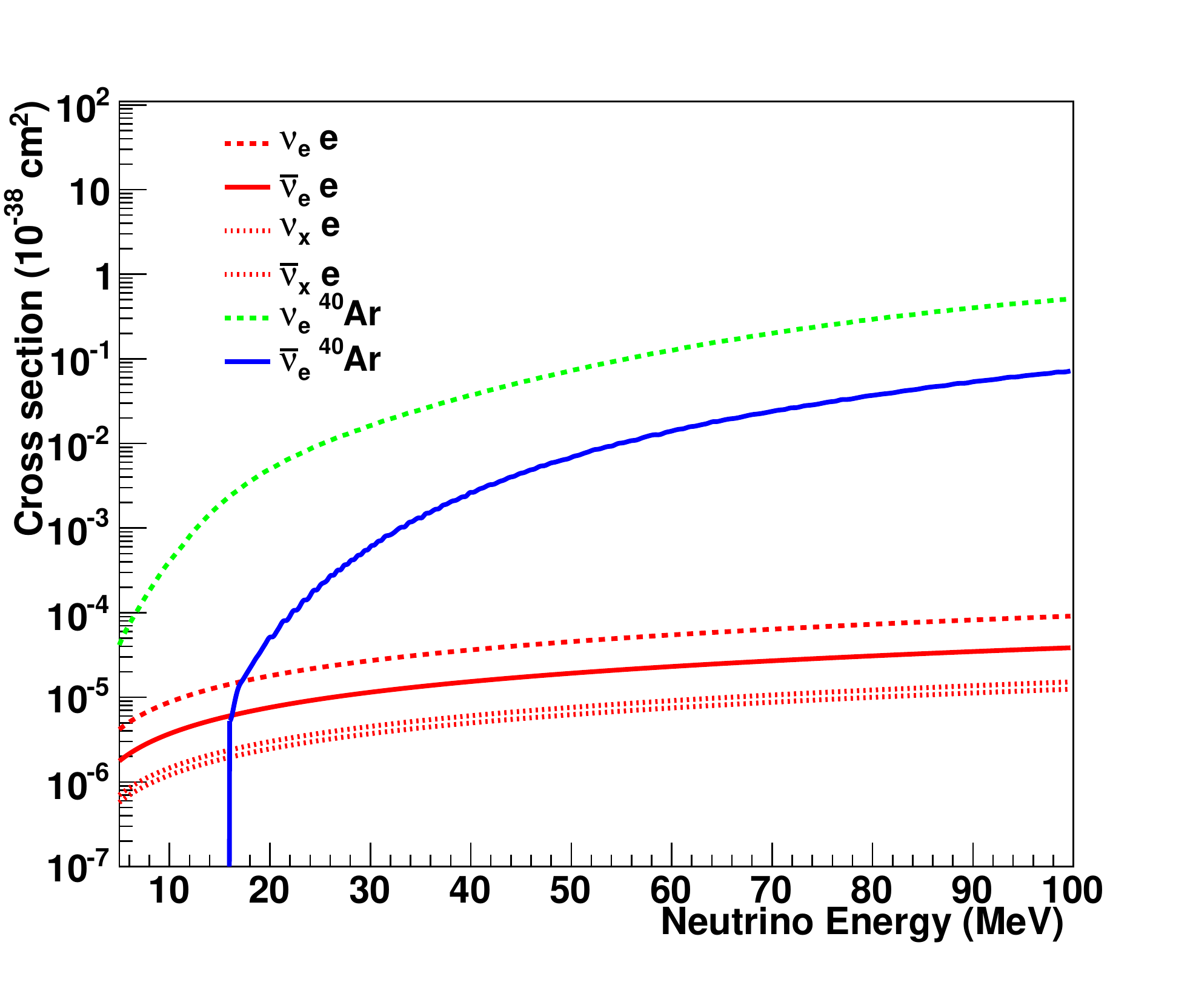}
    \caption{Cross sections for supernova-relevant interactions in argon as a function of neutrino  energy.}
    \label{fig:2}
    \end{figure}

Among the different detection channels in LAr for the neutrino interactions, the dominant interaction is the charged-current (CC) absorption of $\nu_e$ on $^{40}Ar$, for which the observable are short electron tracks plus deexcitation products from the excited $^{40}K^{\ast}$ final state. Then, CC interactions of $\nu_e$ at MeV energies create short electron tracks in liquid argon, potentially accompanied by gamma ray and other secondary particle signatures. Additional channels include a $\bar{\nu}_e$ CC interaction and electron scattering. Cross sections for the most relevant interactions are shown in Fig.~\ref{fig:2}.

The predicted event rate from a supernova burst or the solar neutrinos may be calculated by folding expected neutrino flux differential energy spectra with cross sections for the relevant channels, and with detector response; this is done using SNOwGLoBES~\cite{snowglobes}. 

We use the MARLEY (Model of Argon Reaction Low Energy Yields) generator~\cite{marley} to simulate tens-of-MeV neutrino-nucleus interactions in liquid argon. The LArSoft \cite{larsoft} Geant4-based software package is used to simulate the final-state products from MARLEY in the DUNE LArTPC. SNB and solar neutrino events due to their low energies manifest as spatially small events of few centimeters.

Understanding of cosmogenic and radiological backgrounds, dominated by $^ {39}$Ar, is necessary, although we expect a minor impact on reconstruction, the triggering efficiency for SNB neutrinos could be affected. Background generation follows the \textit{BxDecay0} package\footnote{https://github.com/BxCppDev/bxdecay0}, a C++ library providing simulated nuclear decays that is integrated into LArSoft. A radiological model is being developed for the DUNE FD considering the HD and VD technologies. It includes radioactive decays in the LAr bulk ($^{39}$Ar, $^{42}$Ar, $^{85}$Kr, $^{222}$Rn and their decay chains), the cathode (drifted $^{42}$K from LArSoft's $^{42}$Ar decays, $^{40}$K and the $^{238}$U decay chain), the charge readout planes ($^{60}$Co and the $^{238}$U decay chain), the photon detection system ($^{222}$Rn decay chain), and external sources (gammas and neutrons from surrounding rocks).

\section{Supernova neutrinos in DUNE}

\begin{figure}
\centering
\label{fig:SN-time}
  \includegraphics[width=0.7\textwidth]{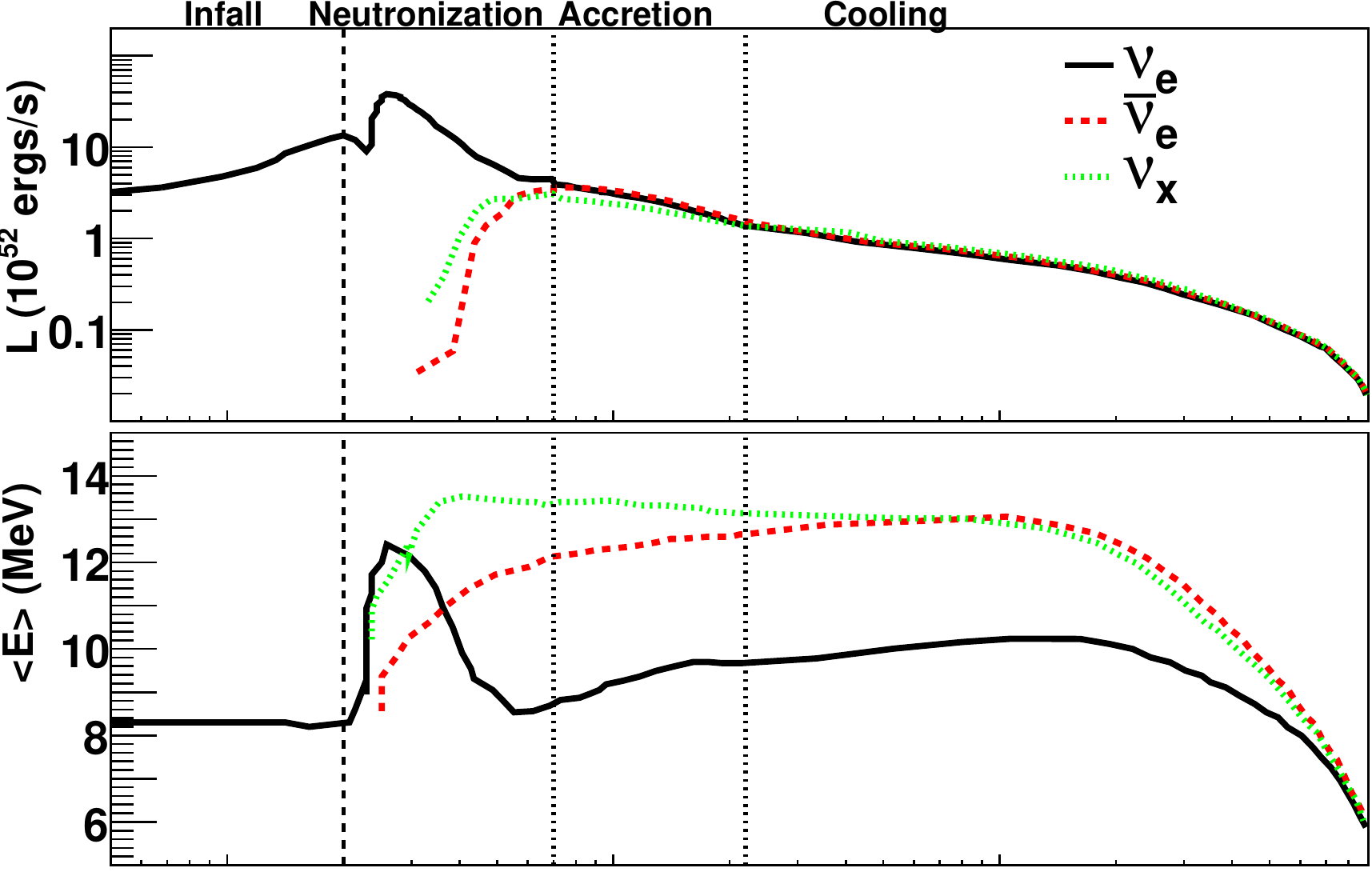}
 \caption{Expected time-dependent flux parameters for a specific model for an electron-capture supernova~\cite{Huedepohl:2009wh}. No flavor transitions are assumed. The top plot shows the luminosity as a function of time, and the bottom plot shows average neutrino energy.}

\end{figure}

Each supernova releases an intense source of neutrinos of all flavors. During a supernova explosion, 99\% of the gravitational binding energy of the star ($\sim10^{53}$~ergs) is released as neutrinos and antineutrinos of all flavors, which play the role of astrophysical messengers, escaping from the SN core. In the event of a galactic supernova explosion, DUNE data will probe the inner evolution of the core-collapse mechanism by studying the time and energy spectra of neutrinos arriving at DUNE. SNB neutrinos are emitted in a burst of a few tens of seconds duration~\cite{Huedepohl:2009wh}. Within DUNE, three qualitative stages of the collapse can be distinguished, as shown in Figure~\ref{fig:SN-time}: 

\begin{enumerate}
    \item The neutronization burst -- a large pulse of $\nu_e$  emission takes place in the first 10's of ms as electrons capture on protons in the stellar core during the formation of a proto-neutron star.  A neutrino sphere forms around the proto-neutron star, inside which the density is so large neutrinos become trapped at which point the neutronization burst of $\nu_e$  emission quenches.
    \item The accretion phase -- during accretion, lasting from tens to hundreds of ms, neutrino emission is dominated by infall of gas from the outer layers of the progenitor onto the outer extant of the proto-neutron star.
    \item The cooling phase -- after infall, the proto-neutron star cools over several seconds.  The neutrino opacity drops, allowing neutrinos to escape the core.  While trapped within the core, neutrino species thermalize so that there is nearly luminosity equipartition between neutrino species during the cooling phase.
\end{enumerate}

The predicted event rate from a SNB is calculated by folding together expected neutrino differential energy spectra, cross sections for the relevant channels, and detector response using SNOwGLoBES. Monte Carlo simulated events are generated using the time and energy of incident neutrinos for a particular SN model using the MARLEY interaction model to simulate the $\nu_e$ CC neutrino interaction and using the standard Geant4-based detector models to simulate the DUNE far detector. Standard LArTPC algorithms are applied to reconstruct electron tracks. All visible energy from the event is used to calculate the incident neutrino energy calorimetrically. Photon detectors are typically used to determine the time of events, but DUNE is exploring how photon detector calorimetry can expand its low-energy physics reach.

In DUNE, the trigger on a SNB can be done using either TPC or photon detection system information. In both cases, the trigger scheme exploits the time coincidence of multiple signals over a timescale matching the supernova luminosity evolution. A redundant and highly efficient triggering scheme is under development. 

A number of astrophysical phenomena associated with supernovae are expected to be observable in the supernova neutrino signal, providing a remarkable  window into the event. In particular, the supernova explosion mechanism, which in the current paradigm involves energy deposition into the stellar envelope via neutrino interactions, is still not well understood, and the neutrinos themselves will bring the insight needed to confirm or refute the paradigm. There are many other examples of astrophysical observables, more details can be found in~\cite{DUNE_SNB}.

Detecting neutrinos from a SNB, we can also learn a lot about neutrinos and particle physics. A SNB can be thought as an extremely hermetic system, which can be used to search for new new physics like Goldstone bosons, neutrino magnetic moments, "dark photons", "unparticles", extra-dimensional gauge bosons, and  sterile neutrinos. Also, self-interactions of neutrinos, neutrino instability, and light gauge bosons can be studied. Such energy-loss-based analysis will make use of two types of information. First, the total energy of the emitted neutrinos compared with the expected release in the gravitational collapse. Second, the rate of cooling should be measured and compared with what is expected from diffusion of the standard neutrinos. As DUNE is mostly sensitive to $\nu_e$, complementary data of $\bar{\nu}_e$ from water Cherenkov and scintillator experiments for careful analysis of the flavor transition will be very useful. The flavor oscillation  neutrino physics and its signatures are a major part of the physics program in the different periods.

\section{Solar neutrinos in DUNE}

\begin{figure}
\label{fig:solar}
\begin{center}
    \includegraphics[width=0.65\linewidth]{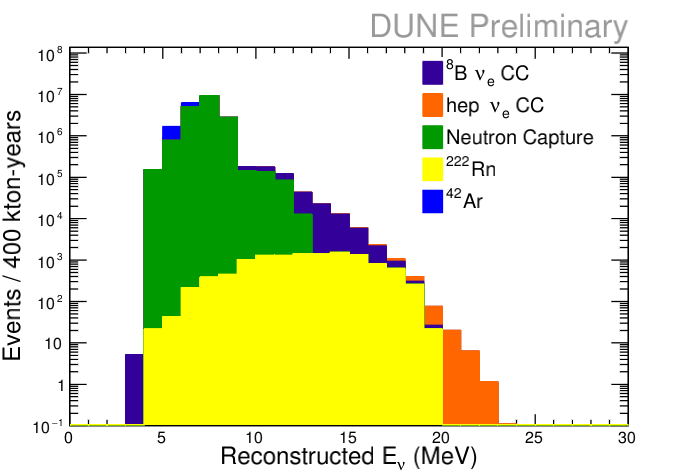}
    \caption{Simulated solar neutrino spectrum with background for the DUNE Far Detector.}
    \end{center}
\end{figure}

Detection of solar and other low-energy neutrinos is challenging in a LArTPC because of relatively high intrinsic detection energy thresholds for the charged-current interaction on argon of about 5\,MeV and because radioactive backgrounds in the same energy regime can affect triggering capability. However, compared with other technologies, a LArTPC offers a large cross section and unique potential channel-tagging signatures from deexcitation photons. Furthermore, observed energy from the final state CC interaction is much more tightly correlated with the incident neutrino energy on an event-by-event basis than the electron recoil spectrum from the ES channel that has been used for past solar neutrino observations such as in Super-Kamiokande~\cite{Super-Kamiokande:2016yck}. Due to this, DUNE will make more precise spectral measurements. Though background rates are large, LArTPC detector allows for background reduction using fiduacialization techniques and the solar neutrino event rate is also substantial in the DUNE far detector, $\sim$100 per day, allowing samples of a few $10^5$ events after 10 years of data collection. Initial studies suggest DUNE could improve the measurement of $\Delta m^2_{21}$ as well as observing the $hep$ and $^8$B solar neutrino flux, as shown in Figure~\ref{fig:solar}.

Detailed studies of solar neutrino detection capability are underway in DUNE along with subsequent physics sensitivity studies. Similarly, DUNE can search for the Diffuse Supernova Neutrino Background (DSNB)~\cite{Beacom:2010kk} at energies just above the endpoint of the solar neutrino spectrum. As DUNE is primarily sensitive to the $\nu_e$ component, DUNE will be the only running experiment with sensitivity to the neutrino component of the DSNB. 

\section{Conclusions}

The DUNE experiment will be sensitive to neutrinos with about 5 MeV up to several tens of MeV, the regime of relevance for core-collapse supernova burst and solar neutrinos. This low-energy regime presents particular challenges for triggering and reconstruction. The combined information from DUNE's TPC and PDS systems will provide a good reconstruction of these events. Software tools that enable preliminary physics and astrophysics sensitivity studies have been developed. The observation of a burst will also enable sensitivity to neutrino mass ordering, and potentially many other topics. In addition, there is discovery potential for $hep$ neutrinos in DUNE and perform a precision measurement of neutrino mixing and fluxes.

\section*{Acknowledgments}

This project has received funding from the European Union Horizon~2020 Research and Innovation programme under Grant no. 101004761, from Spanish Ministerio de Economia y Competitividad (SEIDI-MINECO) under Grant no.~FPA2016-77347-C2-1-P; and from the Comunidad de Madrid.

\bibliographystyle{JHEP}
\bibliography{biblio}

\providecommand{\href}[2]{#2}\begingroup\raggedright\begin{thebibliography}{10}

\bibitem{DUNE_LBL}
{\scshape DUNE} collaboration, B.~Abi et~al., \emph{{Long-baseline neutrino
  oscillation physics potential of the DUNE experiment}},
  \href{http://dx.doi.org/10.1140/epjc/s10052-020-08456-z}{\emph{Eur. Phys. J.
  C} {\bfseries 80} (2020) 978}.

\bibitem{DUNE_SNB}
{\scshape DUNE} collaboration, B.~Abi et~al., \emph{{Supernova Neutrino Burst
  Detection with the Deep Underground Neutrino Experiment}}, {\emph{Eur. Phys.
  J. C} {\bfseries 81} (2021) 423},
  [\href{https://arxiv.org/abs/2008.06647}{{\ttfamily 2008.06647}}].

\bibitem{DUNE_BSM}
{\scshape DUNE} collaboration, B.~Abi et~al., \emph{{Prospects for beyond the
  Standard Model physics searches at the Deep Underground Neutrino
  Experiment}},
  \href{http://dx.doi.org/10.1140/epjc/s10052-021-09007-w}{\emph{Eur. Phys. J.
  C} {\bfseries 81} (2021) 322}.

\bibitem{tdr_v4}
{\scshape DUNE} collaboration, B.~Abi et~al., \emph{{Deep Underground Neutrino
  Experiment (DUNE), Far Detector Technical Design Report, Volume IV Far
  Detector Single-phase Technology}},
  \href{http://dx.doi.org/10.1088/1748-0221/15/08/T08010}{\emph{JINST}
  {\bfseries 15} (2020) T08010},
  [\href{https://arxiv.org/abs/2002.03010}{{\ttfamily 2002.03010}}].

\bibitem{snowglobes}
\emph{{SNOwGLoBES}},
  \href{https://arxiv.org/abs/http://www.phy.duke.edu/~schol/snowglobes}{{\ttfamily
  http://www.phy.duke.edu/~schol/snowglobes}}.

\bibitem{marley}
S.~Gardiner et~al., \emph{{MARLEY (Model of Argon Reaction Low Energy
  Yields)}},  \href{https://arxiv.org/abs/http://www.marleygen.org}{{\ttfamily
  http://www.marleygen.org}}.

\bibitem{larsoft}
\emph{{LArSoft}},  \href{https://arxiv.org/abs/http://larsoft.org}{{\ttfamily
  http://larsoft.org}}.

\bibitem{Huedepohl:2009wh}
L.~Hudepohl, B.~Muller, H.-T. Janka, A.~Marek and G.~Raffelt, \emph{{Neutrino
  Signal of Electron-Capture Supernovae from Core Collapse to Cooling}},
  \href{http://dx.doi.org/10.1103/PhysRevLett.104.251101}{\emph{Phys. Rev.
  Lett.} {\bfseries 104} (2010) 251101},
  [\href{https://arxiv.org/abs/0912.0260}{{\ttfamily 0912.0260}}].

\bibitem{Super-Kamiokande:2016yck}
{\scshape Super-Kamiokande} collaboration, K.~Abe et~al., \emph{{Solar Neutrino
  Measurements in Super-Kamiokande-IV}},
  \href{http://dx.doi.org/10.1103/PhysRevD.94.052010}{\emph{Phys. Rev. D}
  {\bfseries 94} (2016) 052010},
  [\href{https://arxiv.org/abs/1606.07538}{{\ttfamily 1606.07538}}].

\bibitem{Beacom:2010kk}
J.~F. Beacom, \emph{{The Diffuse Supernova Neutrino Background}},
  \href{http://dx.doi.org/10.1146/annurev.nucl.010909.083331}{\emph{Ann. Rev.
  Nucl. Part. Sci.} {\bfseries 60} (2010) 439--462},
  [\href{https://arxiv.org/abs/1004.3311}{{\ttfamily 1004.3311}}].

\end{thebibliography}\endgroup

\end{document}